Develoment of thin high-pressure-laminate RPC electrodes for future high-energy experiments

Kyong Sei Lee[a], Giuseppe Iaselli[b], Youngmin Jo [a], Minho Kang [a], Tae Jeong Kim[a], Dayron Ramos Lopez[b], and Gabriella Pugliese[b]

[a]*Department of Physics and Institute of Basic Sciences, Hanyang University, 42, Majo-ro, Seongdong-gu, Seoul, 04761, Republic of Korea*

[b]*Dipartimento Interateneo di Fisica, Bari Polytechnic, Via Amendola 173, Bari, 70126, Italy*

E-mail: kyong.sei.lee@cern.ch

**Abstract**

In this R&D, an innovative method for producing thin high-pressure laminate (HPL) electrodes for resistive plate chambers (RPC) for future high-energy experiments is introduced. Instead of using thick phenolic HPL (2-mm thick Bakelite), which has been used for conventional RPC triggers, the RPC electrodes in the present study are constructed by bonding 500 μm-thick melamine-based HPL to a graphite-coated polycarbonate plate. A double-gap RPC prototype to demostrate the present technology has been constructed and tested for cosmic muons. Furthermore, the uniform detector characteristrics shown in the test result allows us to explore the present technology in future high-energy experiments.

## 1. Introduction

The necessity of resistive plate chambers (RPCs) for large-area detection with relatively simple detector structure is strong for future high-energy experiments, especially such as Future Circular Collider experiments (FCC-ee/hh) [1-3]. The RPCs whose electrodes are constructed with high-pressure phenolic laminate (HPL) have been satisfactory in terms of triggering performance in previous experiments such as CMS, ATLAS, and ALICE in LHC [4-7]. Their typical time and positional resolutions are about 1 ns and a half cm, respectively. The range of particle response rates is from about 1 to a maximum of a few kHz cm$^{-2}$.

The necessity of R&D efforts of the LHC experiments to obtain better time resolution and higher response rates has been a strong issue for more reliable RPC trigger operations in many high-luminosity collider experiments. To achieve the desired time resolution, the RPC gaps thickness was reduced from 2 mm to 1.4 mm for the improved CMS RPCs [8] and to 1.0 mm for the new ATLAS RPCs [9].

According to the Shockley-Ramo theorem applied to RPCs [10, 11], the thickness of the RPC electrodes should be appropriately adjusted to the gaps thickness. The attenuation factor of the pickup signals induced on the strip plane, $A_{att}$, is $1 + 2d/\varepsilon_r g$, where $d$, $\varepsilon_r$, and $g$ are the thickness, the dielectric constant of the RPC electrodes, and the gaps thickness, respectively. Considering the fixed value of $\varepsilon_r$ (~ 3.5) for typical HPL, the ideal thickness $d$ is obviously smaller than the value of $g$. However, the gaps merely made of thin HPL for RPC electrodes are mechanically unstable because of the high flexibility of the material, which would make it difficult to guarantee the mechanical homogeneity required for consistent efficiency and response time over the whole detector region.

An innovative way to avoid the mechanical instability of thin RPC gaps can be achieved by providing additional mechanical support to the thin HPL. Each RPC electrode is constructed by painting with semiconductive graphite and then by gluing a 0.5-mm thick HPL layer onto a 1-mm thick polycarbonate (PC) panel. The thicknesses of HPL and PC panel are the same for the cathode- and anode side electrodes. The thickness of RPC gaps maintained by the circular spacers is 1.0 mm.

The choice of thinner thickness for HPL is also beneficial for improving the rate capability of the detectors. The rate capability would be inversely proportional to the electrode thickness and the averaged avalanche charge [11]. However, considering the increase in capacitance formed by the electrodes and gaps with decreasing thicknesses, $d$ and $g$, the RC time constant of charging RPC gaps [12] remains about the same despite decreasing $d$. Therefore, the enhancement in the rate capability is expected to be less sensitive to the reduction ratio of $d$.

Considering the difficulty of developing a new low-resistance RPC electrode material at a reasonable cost, the use of digitization electronics with a better sensitivity is more efficient to improve the rate capability.

Nevertheless, thinner electrodes and gaps are still advantageous to suppress the lateral spread of the pick-up signal induction on the strips as well as to improve the intrinsic time resolution of the RPC detector.

## 2. Construction of a prototype detector

### 2.1. Mechanical structure

The structure of the thin HPL double-gap RPC of the present R&D is illustrated in Fig. 1. To provide conductivity to the RPC electrodes, semi-conductive graphite resin is painted on 1-mm polycarbonate panels. The surface resistivity of the graphite resin is adjusted by mixing conductive ($\rho \sim 1$ $\Omega$cm) and black carbon ($\rho \sim$ a few k$\Omega$cm) resins in appropriate proportions. The measured surface resistivity values for the cathode- and anode side electrodes are about 50 and 400 k$\Omega$/□, respectively.

The active area of the double-gap RPC prototype is $96 \times 56$ cm$^2$. The pitch and length of the 28 copper strips shown in the upper figure of Fig. 2 are 20 and 103 cm, respectively. A 32-channel front-end electronics (FEE) board is placed close on each side of the detector to read the detector pulses traveling toward both opposite ends of the strips, as shown in the lower figure of Fig. 2. The detector pulses are transferred to the FEE boards by 10-cm long 50-$\Omega$ coaxial cables.

The uniform travelling speed of the detector pulses in the 20 mm spacing strips with a transmission impedance of about 20 $\Omega$ (18.4 cm/ns, measured by function generator pulses) allows us to determine the position of a particle hit in the strip direction by measuring the time difference between the pulses arriving at the two opposite strips ends.

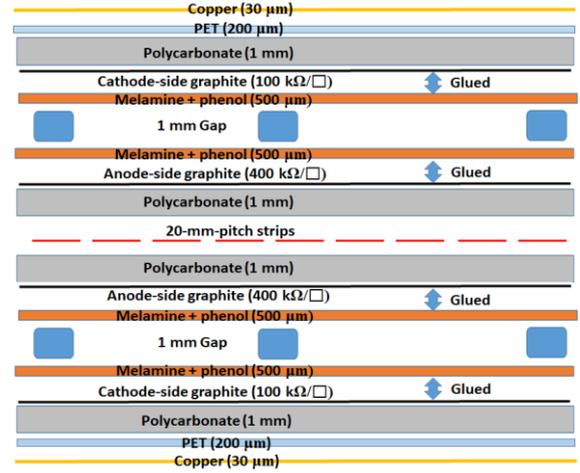

**Fig. 1**. Schematic diagram of the double-gap RPC prototype.

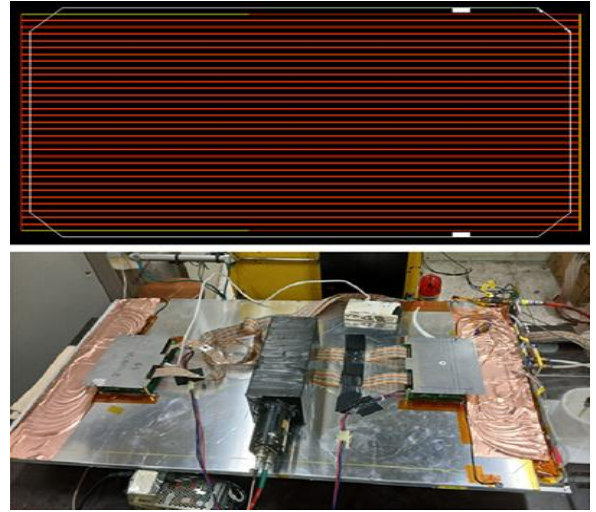

**Fig. 2.** 28 strips with a 20-cm pitch placed between two RPC gaps (top) and a prototype detector equipped with two 32-channel FEE boards (bottom).

An additional 200-μm thick PET film is placed on the 1-mm thick PC at the electrode on the cathode side of each gap to reinforce the electrical insulation of the working high voltage from the copper ground that surrounds the double-gap structure. The copper ground also provides a reference ground for the FEE boards.

Because PC is excellent for electrical insulation,

it is not neccessary to glue the PET film to the PC on the cathode side with an insulating adhesive such as ethylene vinyl acetate (EVA). The relatively hygroscopic HPL layers in the electrodes are closely insulated by 1-mm thick PC and 15-mm wide 'T'-shaped PC bars, which are inserted along the periphery of the gap to provide a gas seal. The 1 mm thick T-shaped bars maintain the 1 mm thick gaps.

## 2.2. Thin HPL for RPC electrodes

HPL, the resistive part of RPC electrodes, is commercially used for typical home furniture. The HPL in this study is composed of double layers of 200-μm-thick glossy melamine and 300 μm-thick matt-finish phenol. Each HPL sheet is directly bonded to a graphite-painted PC panel using epoxy. The influence of a few μm thick epoxy ($\rho \sim 10^{12}$ Ωcm) on the effective electrode resistivity of is expected to be insignificant.

Four HPL samples were placed in an isothermal box made of 1 cm thick plexiglass (polymethyl methyacrylate) to monitor and measure their resistivity under various temperature and humidity conditions. Figure 3 shows the resistivity of four samples labeled S1, S2, S3, and S4, calibrated to a standard temperature of 20 °C, as a function of humidity. The exponential function, $\rho(T) = \rho(20) \times e^{-\alpha(T-20)}$, was applied for temperature calibration. Here, the temperature coefficient, $\alpha$, was set to 0.12/°C. As shown in Fig. 3, the average resistivity of the four samples, calibrated to $T = 20$ °C and measured at 50% humidity, is about $1.5 \times 10^{11}$ Ωcm. Each 10% increase in humidity reduces the resistivity of HPL by about 30%.

## 2.3. Varnishing of gas gaps with linseed oil

The inner surfaces of the RPC gaps are varnished with a mixture of 50% linseed oil and 50% heptane (volume ratio) and the same procedure that were used for the improved RPCs in the CMS experiment [5, 8].

The temperature of the linseed-oil mixture is maintained at 28 °C at which the optimal oil viscosity for the varnishing procedure is achieved with the oil and heptane mixture ratio. The oil mixture is inserted and removed through the lower corner gas inlets of the gas gaps.

To dry the oil layers, filtered air supplied by a compressor is immediately introduced through the same lower corner gas inlets of the gas gaps, right after the oil varnishing procedure is completed. The air flow rate fed to the RPC gap and the duration are about 100 $l/h$ and 6 days, respectively. During the dry procedure (polymerization of oil layers), the RPC gaps and the oil mixture are maintained at 30 °C. The relative humidity of the air is properly adjusted to the ambient humidity in the laboratory. The minimum to avoid vending of the RPC gaps during the dry seasons is 30% while 40% is the maximum in summer to guarantee the high rigidity of the dried oil layers.

It was found that the oil-varnishing procedure of the thin HPL RPC gaps in the present study is necessary to reduce the stochastic noise of the detector. After the proper oil varnishing for the RPC gaps, the noise rate measured at 20 °C with a digitization threshold of 450 μV decreases from ~ 10 to about 0.5 Hz cm$^{-2}$.

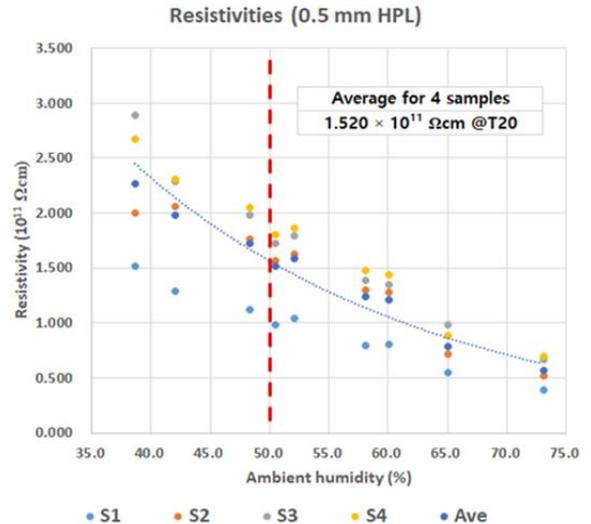

**Fig. 3**. Resistivity for four samples labelled S1, S2, S3, and S4, calibrated for a standard temperature of 20 °C, as a function of humidity.

## 3. Tests

### 3.1 Electronics and triggers

The thin HPL double-gap RPC was tested for cosmic muons using a 64-channel multi-hit Time-to-Digital-Converter (TDC) based on VME. The root-mean-squared (RMS) time resolution of each TDC channel is about 700 ps. Therefore, the TDC RPC data measured for cosmic muons and stochastic noise are digitized with a resolution of 1-ns.

The cosmic muons are tagged by two plastic scintillators of size $10 \times 10 \times 20$ cm$^3$. They are installed closely between the RPC prototype as shown in the lower figure of Fig. 2. The estimated time resolution for the muon triggers provided by two plastic scintillators with an overlap length of 20-cm is about 300 ps.

The 32-channel front-end electronics (FEE) boards used to digitize the RPC pulses are customized electronics that have been utilized for basic R&D works to improve the CMS RPCs [13] and for development of multi-gap RPCs for photon nondestructive imaging [14]. The RPC signals are fed into the FEE boards with an input impedance of 20 Ω and are linearly amplified by a gain of 200 before digitization. The digitization threshold for the signals travelling toward the both ends of strips is set to 450 μV which is roughly equivalent to 60 fC in charge sensitive-mode electronics.

### 3.2. Gas mixture for detector tests

The gas mixture for RPC operation consists of a standard tetrafluoroethane (TFE)-based gas mixture (95.2% $C_2H_2F_4$, 4.5% i-$C_4H_{10}$, and 0.3% $SF_6$) that has been used to operate phenolic RPCs in many large-scale RPC systems, including the CMS and ATLAS of the current LHC experiment [4-8].

The efficiencies and average cluster sizes measured near the center of the detector are shown in Fig. 4. The effective high voltages, $HV_{eff}$, corrected for pressure and temperature are values independent of environmental conditions. The efficiencies and average cluster sizes are scaled on the left vertical axis and on the right vertical axis, respectively. Data measured at the left- and the right-side ends of the strips are labelled as open and solid circles, respectively.

The solid line in Fig. 4 was obtained by parametrizing the coincident efficiencies of the left- and the right-side data using the sigmodal function [15] of the $HV_{eff}$. Then, the working-point high voltage, $HV_{WP}$, for the coincident measurement is defined as $HV_{0.95} + 0.15$ kV. Here, $HV_{0.95}$ is the value of $HV_{eff}$ that gives 95% efficiency with respect to the maximum efficiency predicted by fitting the sigmoid function. The dashed curves parameterized by polynomial functions merely indicate increasing trends in the average cluster sizes with the $HV_{eff}$.

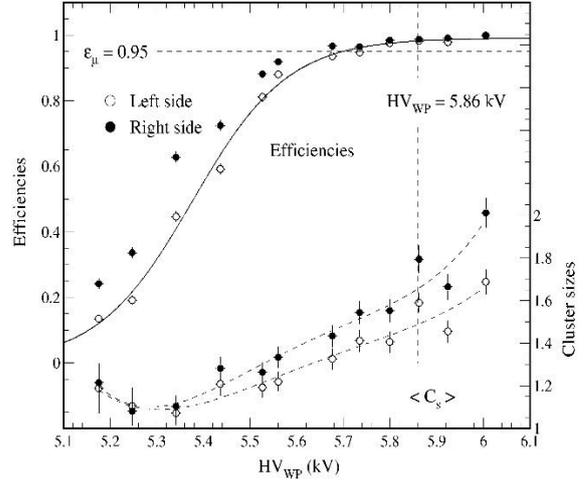

**Fig. 4**. Efficiencies and average cluster sizes measured at a threshold of 450-μV. The notation details are described in the text.

The difference of about 100 V in $HV_{eff}$ between the left (open circles) and the right (solid circles) efficincies is attributed to the small difference in the amplification gain of the two FEE boards, which are prepared for the test without precise calibration. The smaller cluster sizes for the right-side strip data are also attributed to the same reason. The coincidence of efficiency and average cluster size at the $HV_{WP} = 5.86$ kV is estimated to be 0.981 and about 1.7, respectively.

Figure 5 shows the channel distribution of the stochastic noises measured at the left- and the right-sided strip ends at $HV_{WP} = 5.9$ kV for 6.5 s in the left and the right figures, respectively. The estimated noise hit rate is 0.36 Hz cm$^{-2}$.

The uniformity of the detector response is tested also at $HV_{eff} = 5.9$ kV. The efficiencies and average cluster sizes are measured near the center and at three different locations near each strip end of the detector. The efficiencies and average cluster sizes for the data measured from the left- ($\varepsilon_L$ and $<C_{sL}>$) and right-side ($\varepsilon_R$ and $<C_{sR}>$ strip ends, and the left-right coincidence of efficiencies ($\varepsilon_C$) are listed in the Table 1.

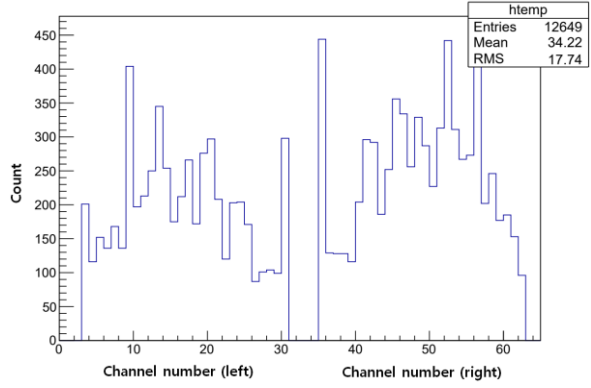

**Fig. 5**. Channel distributions of the stochastic noises measured at the left- (left) and right-sided (right) strip ends for 6.5 s at $HV_{WP}$ = 5.9 kV.

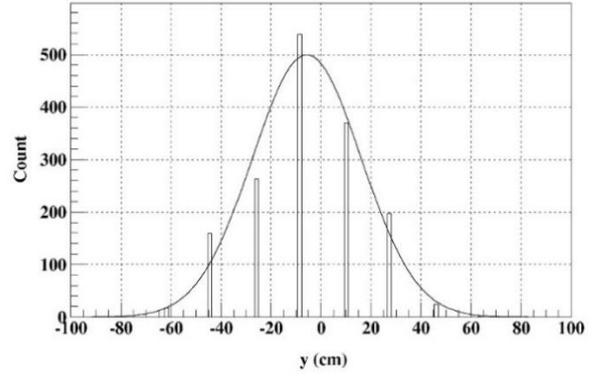

**Fig. 6.** Hit positions in the strip direction determined by the time difference in the TDC data from the left and the right ends of the fastest strip.

| Places | $\varepsilon_L$ | $<C_{sL}>$ | $\varepsilon_R$ | $<C_{sR}>$ | $\varepsilon_C$ |
|---|---|---|---|---|---|
| Center | 0.995 | 1.98 | 0.988 | 1.73 | 0.988 |
| Up-left corner | 0.975 | 1.84 | 0.960 | 1.73 | 0.960 |
| Middle left | 0.990 | 2.24 | 0.980 | 1.98 | 0.980 |
| Down-left corner | 0.972 | 1.90 | 0.958 | 1.69 | 0.958 |
| Up-right corner | 0.976 | 1.76 | 0.975 | 1.70 | 0.975 |
| Middle right | 0.990 | 2.07 | 0.985 | 1.96 | 0.985 |
| Down-right corner | 0.988 | 1.95 | 0.975 | 1.78 | 0.975 |

**Table 1**. Efficiencies and average cluster sizes for the data measured from the left- ($\varepsilon_L$ and $<C_{sL}>$) and right-side ($\varepsilon_R$ and $<C_{sR}>$ strip ends, and the left-right coincidence of efficiencies ($\varepsilon_C$).

The relatively low coincidence efficiencies measured in the upper left and the lower left corners can be attributed to the partial overlap of the muon tagging regions with graphite displacement regions lying along the gap peripheries.

The present RPC prototype is designed to measure the particle hit position also in the strip direction. Applying the pulse-travelling speed of 18.4 cm/ns in 20 mm pitch strips, the Gaussian-fit time resolution for the hit position, determined by the time difference in the fastest TDC data from the left and right ends of the fastest strip, is about 22 cm as shown in Fig. 6.

The relatively poor time resolution of the FEEs (~ 300 ps) and TDC (~700 ps) account for the wide distribution in hit positions ('$y$' in the figure). To determine the hit positions with a resolution of less than 2 cm, the time accuracy for both FEEs and trigger electronics should be better than 50 ps, which is also required to achieve the time resolution of about 300 ps expected for the 1-mm gap-thickness RPC.

## 4. Summary

We have constructed a thin HPL double-gap RPC module and tested the detector performance with the typical gas mixture for RPC operation in the LHC experiments. The summary of the present results is as follows:

(1) The first prototype of the thin HPL double-gap RPC was successfully constructed with the 500-μm thick HPL. The reliable detector construction procedure was well achieved.

(2) The performance of the prototype detector tested for cosmic muons was fairly stable over time with a relatively low stochastic noise hit rate of about 0.4 Hz cm$^{-2}$.

(3) The coincident efficiency of the RPC measured at $HV_{WP}$ = 5.86 kV was 0.981 with an average cluster size of aproximately 1.7. The uniformity of the detector characteristics in the detection areas was also well confirmed.

(4) The development of new FEE with a time resolution better than 50 ps and tests with a higher TDC resolution are the next phase of research to obtain the detector characteristics that are the subject of this study.

**Acknowledgments**

This study was supported by the National Research Foundation of Korea (grant number: RS-2024-00337656).